\documentclass[aps, pre,twocolumn,superscriptaddress]{revtex4}
\usepackage{epsfig,amssymb,amsmath,amsthm,amsfonts,amsbsy,mathrsfs,bm}
\usepackage{graphicx}
\usepackage{color}
\usepackage{siunitx}
\usepackage{bm}

\usepackage{tikz}
\usetikzlibrary{arrows,shapes,chains}
\newcommand*\circled[1]{\tikz[baseline=(char.base)]{
\node[shape=circle,draw,inner sep=0.5pt] (char) {#1};}}

\begin{document}
\title{Phase behaviour of Lennard-Jones particles in two dimensions}
\author{Yan-Wei Li}
\affiliation{Division of Physics and Applied Physics, School of Physical and
Mathematical Sciences, Nanyang Technological University, Singapore 637371,
Singapore}
\author{Massimo Pica Ciamarra}
\email{massimo@ntu.edu.sg}
\affiliation{Division of Physics and Applied Physics, School of Physical and
Mathematical Sciences, Nanyang Technological University, Singapore 637371,
Singapore}
\affiliation{
CNR--SPIN, Dipartimento di Scienze Fisiche,
Universit\`a di Napoli Federico II, I-80126, Napoli, Italy
}
\date{\today}

\begin{abstract}
The phase diagram of the prototypical two-dimensional Lennard-Jones system, while extensively investigated, is still debated. 
In particular, there are controversial results in the literature as concern the existence of the hexatic phase and the melting scenario. 
Here, we study the phase behaviour of 2D LJ particles via large-scale numerical simulations. 
We demonstrate that at high temperature, when the attraction in the potential plays a minor role, melting occurs via a continuous solid-hexatic transition followed by a first-order hexatic-fluid transition. 
As the temperature decreases, the density range where the hexatic phase occurs shrinks so that at low-temperature melting occurs via a first-order liquid-solid transition. The temperature where the hexatic phase disappears is well above the liquid-gas critical temperature.
The evolution of the density of topological defects confirms this scenario.
\end{abstract}

\maketitle
\section{Introduction}
The first experimental investigations of the melting transition of two-dimensional (2D) solids focused on rare gases adsorbed on graphite~\cite{Brinkman1982, Heiney, Strandburg}. 
Since the Lennard-Jones (LJ) potential well describes the interaction potential between rare gases, these earlier studies triggered several numerical investigations of the phase diagram of two-dimensional LJ systems.
These works followed through the years up to recent times~\cite{LJpd1981,LJ_vapor_liq,Singh90,LJpd_Feng2000, LJpd_Phillips,Hayato_EPL,Wierschem2011,Hajibabaei2019}. 
Debates in the literature concerned the existence of the hexatic phase and the order of the transitions separating the hexatic and the liquid phase.
One possibility is that, in LJ systems, melting occurs through a continuous solid-hexatic transition driven by the unbinding of dislocation pairs followed by a continuous hexatic-liquid one, driven by the further unbinding of isolated dislocation into disclinations, as in the celebrated Kosterlitz-Thouless-Halperin-Nelson-Young (KTNHY) theory~\cite{KT, HN, Y}. 
Alternatively, melting may follow the mixed or hard-disks~\cite{Krauth2011, Krauth2015, OurPaper, Glotzer, ningxu, John_Russo,Experiment_harddisc} scenario, in which the solid-hexatic transition is continuous, and the hexatic-liquid is discontinuous.
Melting might also be discontinuous altogether, in which case there is no hexatic phase.
These possible scenarios occur in inverse power-law repulsive systems, where the stiffness of the interaction potential fixes the melting path~\cite{Krauth2015}, and in hard-polygons, where the melting scenario is controlled by the number of edges~\cite{Glotzer}.
Besides, it is also possible that in LJ systems, and more generally in the presence of attractive forces, different melting scenarios occur in distinct regions of the phase diagram~\cite{Massimo_2020PRLmelting}, given that both density and temperature changes induce melting.
\begin{figure}[tb]
 \centering
 \includegraphics[angle=0,width=0.5\textwidth]{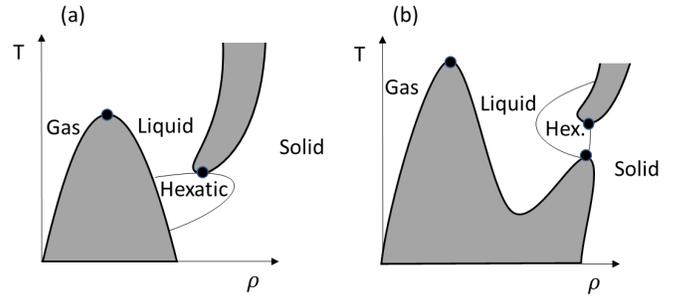}
 \caption{Speculated phase diagrams of the two-dimensional LJ systems, in $T$-$\rho$ plane~\cite{Nelson_book}. The shadowed regions indicate phase coexistence.
\label{fig:pdNelson}
}
\end{figure}

The controversy concerning the features of the melting transition of 2D LJ solids is well illustrated by referring to two possible phase diagrams mentioned in the literature~\cite{Nelson_book}, we reproduce in Fig.~\ref{fig:pdNelson}. 
In both phase diagrams, at very high temperature, the hexatic is absent, and the solid-liquid transition is discontinuous. 
Similarly, no hexatic phase occurs at low temperature, where one observes gas-solid coexistence.
The hexatic phase is bounded by the gas-liquid coexistence curve, in panel (a), or instead may interfere with the solid-liquid coexistence region as in panel (b). 
These are, we remark, just two of many possible~\cite{Nelson1979, Nelson_book} phase diagrams LJ systems might follow.
Indeed, it is now well established that the hexatic region is not bounded at high temperature, as in this limit the attractive tail of the intermolecular potential is negligible, and LJ systems behave as Weeks-Chandler-Anderson (WCA) system~\cite{WCA}, whose interaction potential is obtained by truncated the LJ potential in its minimum, or equivalently as a 
$1/r^{12}$ one. Recent works~\cite{Krauth2015,wcaMelt} have demonstrated that in these systems, the hexatic phase is present.
We note, however, that while these previous work reported a mixed order transition~\cite{Krauth2015,wcaMelt}, at high temperature, a recent investigation of the LJ phase diagram suggested the KTHNY one~\cite{Hajibabaei2019}.
While the existence of the hexatic phase at high temperature is well established, this is not at low temperatures, when the attraction plays a role. Indeed, while some recent works support its existence~\cite{Wierschem2011, Hayato_EPL}, others do not~\cite{Di_SoftMatter, Bo_Prx}.

\begin{figure*}[!!t]
 \centering
 \includegraphics[angle=0,width=0.9\textwidth]{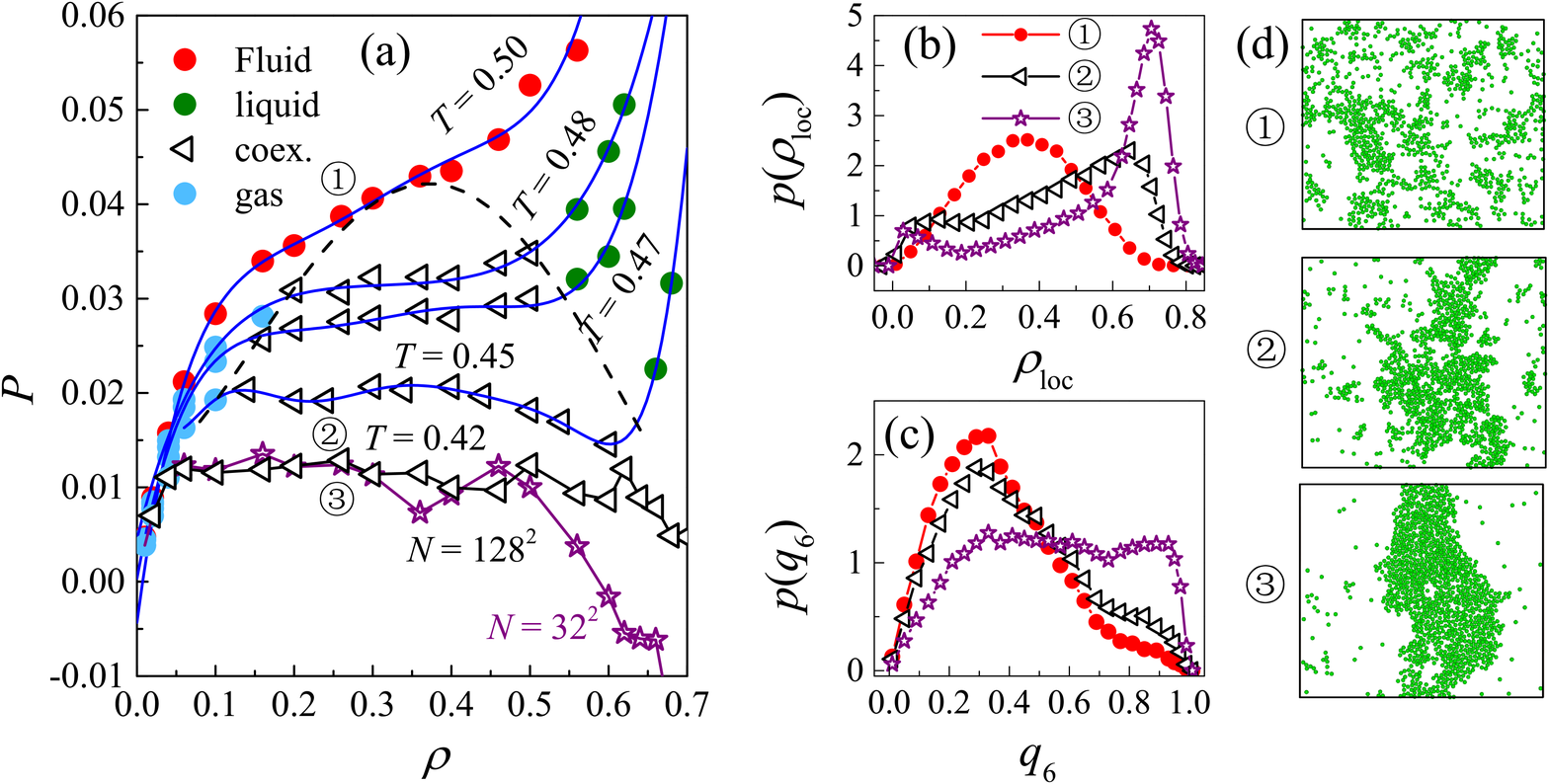}
 \caption{Condensation transition. Panel (a) illustrates the equation of state at different temperatures. 
Different symbols correspond to different phases, as in the legend. 
Blue lines are polynomial fits to the equation of state, while the black and the purple lines connecting at $T=0.42$ are guides to the eye. 
The black dashed line is a polynomial fit of the coexistence boundary, according to which the critical point is at $(\rho_c\simeq0.366, P_c\simeq0.423)$.
Panels (b) and (c) illustrate the probability distribution of the local density and that of the local bond-orientational order parameter at selected state points, as indicated in (a). Panel (d) illustrates snapshots of the system. 
\label{fig:lg}
}
\end{figure*}

\section{Model and methods~\label{sec:technichs}}
\subsection{Simulation details~\label{sec:numerics}}
We study the phase behavior of 2D monodisperse LJ particles of mass $m$, interacting with potential
\begin{equation}
U(r)=
\begin{cases}
4\epsilon[(\sigma/r)^{12}-(\sigma/r)^{6}+C]& r\leq r_c \\
0& \text{otherwise},
\end{cases}
\label{eq:lj}
\end{equation}
where $r_c = 2.5\sigma$, $C$ is a constant chosen such that $U(r_c)=0$.
$\sigma$, $m$ and $\sqrt{m\sigma^{2}/\epsilon}$ will be our units of length, mass and time, respectively.
We mainly consider a system with $N=318^{2}$ particles in a rectangular box with side length ratio $L_{x}:L_{y}=2:\sqrt{3}$. 
Different system sizes ranging from $N=32^{2}$ to $N=512^{2}$ are additionally studied to investigate finite size effects.

We equilibrate and sample the system in the canonical ensemble via molecular dynamics simulation. 
The equations of motion are integrated via a Verlet algorithm~\cite{Allen_book}, and the temperature is fixed via the Nos\'{e}-Hoover thermostat~\cite{Allen_book}. 
We perform the simulations with the GPU-accelerated GALAMOST package~\cite{Galamost}.

\subsection{Identification of the different phases~\label{sec:methods}}
The spatial decay of the translational and bond-orientational correlation functions allow distinguishing the solid, hexatic and liquid phases.
The translational correlation function is $c(r=|\vec{r}_{i}-\vec{r}_{j}|)=e^{i {\vec G} \cdot (\vec{r}_{i}-\vec{r}_{j})}$, where $\vec{G}$ is one of the first Bragg peaks, identified by the static structure factor~\cite{Krauth2011, Massimo_PRE2019}.
The  bond-orientational correlation function is $g_{6}(r=|\vec{r}_{i}-\vec{r}_{j}|)=\langle\psi_{6}(\vec{r}_{i})\psi_{6}^{*}(\vec{r}_{j})\rangle$, where $\psi_{6}(\vec{r}_i)$ is the bond-orientational order parameter of particle $i$, defined as $\psi_{6}(\vec{r}_i)=\frac{1}{n}\sum_{m=1}^{n}\exp(i6\theta_{m}^{i})$. 
Here, $n$ is the number of nearest neighbors of the particle and $\theta_{m}^{i}$ is the angle between $(\vec{r}_{m}-\vec{r}_{i})$ and a fixed arbitrary axis.

In the solid phase, $c(r) \propto r^{-\eta}$ with $\eta \leqslant 1 / 3$, corresponding to the quasi-long-range translational order, a consequence of the Mermin-Wagner theorem~\cite{Mermin}, while $g_{6}(r)$ exhibits almost no decay due to the long-range bond-orientational order. In the hexatic phase, there is an exponential decay in $c(r)$, i.e., $c(r) \propto \exp (-r / \xi)$ and a power-law decay in $g_{6}(r)$, $g_{6}(r) \propto r^{-\eta_{6}}$ with $0<\eta_{6} \leqslant 1 / 4$, representing a short-range translational order and a quasi-long-range bond orientational order. 
In the liquid phase, both the translational and the bond-orientational orders are short-range, resulting in exponential decay in both $c(r)$ and $g_{6}(r)$.

The coexistence phase is identified by the Mayer-Wood loop in the equation of state and the bimodal probability distribution of the local density. 
The Mayer-Wood loop results from the interfacial free energy between coexistence phases and thus is a signature of the first-order transition. 
The coexistence phase is within this loop with the phase boundary determined via the Maxwell construction. 
The local density for each particle is defined as $\rho(\vec{r}_i)=\frac{\sum_{j=1}^{N}H(r_l-|\vec{r}_{i}-\vec{r}_{j}|)}{\pi r_l^2}$, where $H$ is the Heaviside step function, $r_l=50$ for the system with $N=318^{2}$ and $5$ for the system with $N=32^2$. 
The choice of $r_l$ does not affect these results unless $r_l$ becomes tiny or of the order of the system size.

\section{Results \label{sec:results}}
\subsection{Gas-liquid transition \label{sec:cp}}
We begin by briefly reviewing our results for the gas-liquid transition.
We illustrate in Fig.~\ref{fig:lg}(a) the pressure dependence on the density and demonstrate in Figs.~\ref{fig:lg}(b) and \ref{fig:lg}(c) the probability distribution of the local density and of the local bond-orientational order parameter, at $\rho\simeq 0.25$, and at temperatures $T=0.5$, $T=0.45$, and $T=0.42$, for a system of $N=1024$ particles. 
Snapshot of the systems are in Fig.~\ref{fig:lg}(d).
The coexistence region is clearly revealed by the almost density independence of the pressure, the concomitant bimodal density distribution, and the emergence of heterogeneous distribution of particles in the snapshot. 
We determine the phase boundaries via the Maxwell construction and use a polynomial fit of these points to approximate the coexistence curve close to the critical point, which we estimate to occur at $(\rho_c\simeq0.366, P_c\simeq0.423)$. 
Previous studies found comparable results~\cite{LJpd1981,LJ_vapor_liq,LJpd_Feng2000, LJpd_Phillips,Schrader}.

In the $N = 1024$ system, at low temperature $T=0.42$, the pressure is negative at large enough density, as apparent in Fig.~\ref{fig:lg}(a). 
This is a finite size effect, as pressure stays positive as thermodynamics dictates, for $N=128^2$ (black triangles at $T=0.42$ in Fig.~\ref{fig:lg}(a)). 
This system size-dependent behaviour is not expected in the liquid or gas-liquid coexistence region, as these phases have small correlation lengths. 
Instead, this result indicates that $T=0.42$ is below the triple point and that the coexisting phases are of gas and solid type. 
This claim is supported by the probability distribution of the local bond-orientational order parameter $p(q_6)$, where we observe at $T=0.42$ a pronounced peak at $q_6>0.7$ corresponding to well-ordered particles. This well ordered region is absent at higher temperatures, e.g., at $T=0.50$ and $T=0.45$. The temperature of the critical point is, therefore, in the range $T=[0.42:0.45]$.

\subsection{Temperature dependence of the two-dimensional melting \label{sec:melt}}
\subsubsection{Phase determination}
\begin{figure*}[!t]
 \centering
 \includegraphics[angle=0,width=0.9\textwidth]{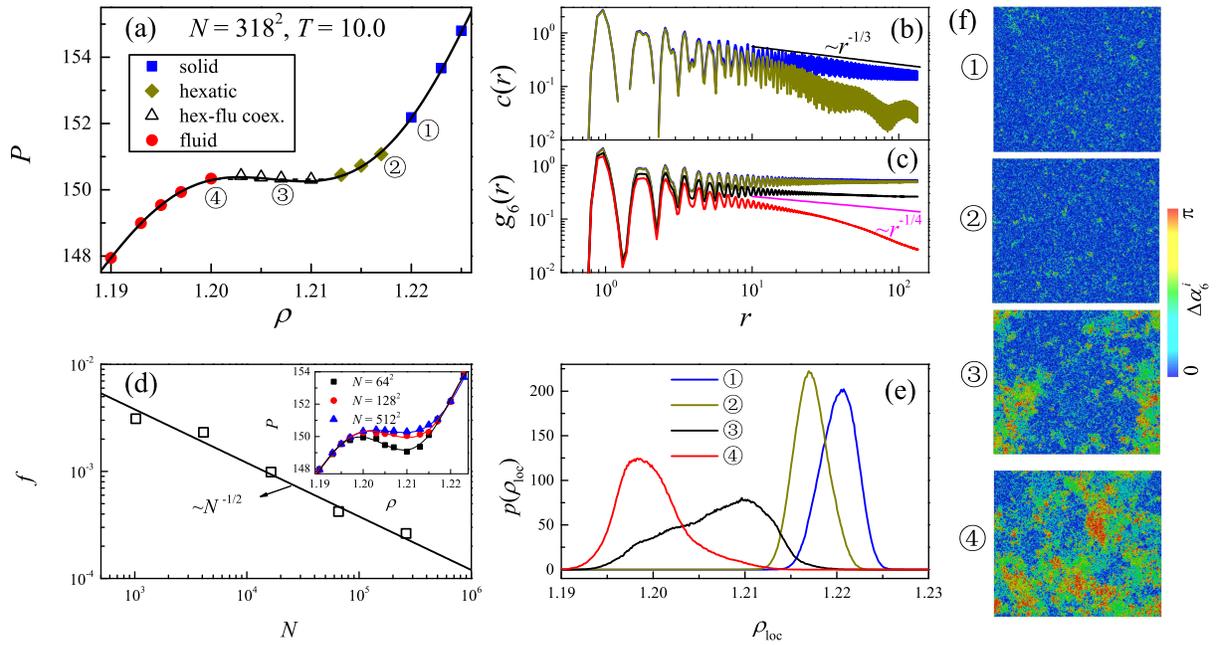}
 \caption{
Melting transition at $T=10.0$.
Panel (a) illustrates the equation of state for a system with $N=318^2$ particles. Different symbols indicate the different phases, as shown in the legend. The black line is a fifth-order polynomial fit. Panels (b) and (c) illustrate the translational correlation function $c(r)$ and the bond-orientational one $g_{6}(r)$ at different densities, as indicated in (a). 
Panel (d) illustrates the system size dependence of the interfacial free energy $f$ and of the equation of state (inset).
Panel (e) shows the distribution of the local density at selected state points, as indicated in (a). 
Panel (f) illustrates snapshots of the system with particles color coded according to the angle $\Delta \alpha^i_4$ between their local bond-orientational order parameter, and the global one.
\label{fig:highT}
}
\end{figure*}

\begin{figure*}[!!t]
 \centering
 \includegraphics[angle=0,width=0.95\textwidth]{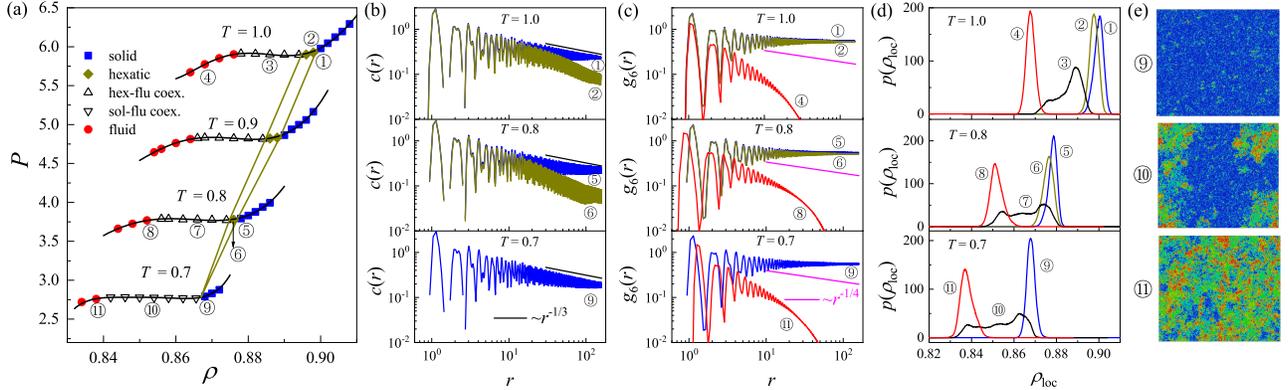}
 \caption{Temperature dependence of the melting transition.
Panel (a) illustrates the equation of state at several temperatures, with different symbols corresponding to different phases, as in the legend. Black lines are polynomial fits.
Panels (b) and (c) illustrate the translational and the bond-orientational correlation functions at selected state points, as indicated in (a). 
Panel (d) show the probability distribution of the local density, and panel (e) snapshot of the system coloured as in Fig.~\ref{fig:highT}(f).
\label{fig:lowT}
}
\end{figure*}

We investigate the melting transition of 2D LJ solids first focusing on a high-temperature value, $T = 10.0$, where attractive forces are negligible.
At this value of the temperature, we observe a clear Mayer-Wood loop in the equation of state, indicating phase coexistence, as in Fig~\ref{fig:highT}(a).
We use Maxwell's construction to locate the phase boundaries, having approximated the equation of state via a fifth-order polynomial. 
The spatial decay of the translational and bond-orientational correlation functions then allows identifying the coexisting pure phases.
The behavior of this quantities is illustrated in Figs.~\ref{fig:highT}(b) and \ref{fig:highT}(c).
Both correlation functions decay exponentially in the low-density coexisting phase, which is therefore of liquid type.
In the high-density coexisting phase $c(r)$ decays exponentially while $g_{6}(r)$ decays as a power-law with the exponent $\eta_{6}<1/4$.
Accordingly, the coexisting phases are of fluid and hexatic type.
The system size scaling of the interfacial free energy $f$ determined from the area covered by the up bump of the Mayer-Wood loop and the horizontal line from the Maxwell construction further supports the first-order nature of the hexatic-fluid transition.
We find $f\propto N^{-1/2}$, as illustrated in Fig.~\ref{fig:highT}(c), providing robust evidence for a discontinuous hexatic-fluid transition at $T=10.0$~\cite{Schrader,Krauth2011}. 
Consistently with this result, the probability distribution of the local density $p(\rho_{\rm loc})$ has a unimodal Gaussian-like shape in the pure phases; Conversely, it is broad, almost bimodal, in the coexistence region.
As the density increases further, the system continuously transients into the solid phase. Here, $c(r)$ decays algebraically  with the exponent $\eta \leqslant 1/3$, while $g_6(r)$ does not decay.

We illustrate the different states by color-coding each particle according to the angle $\Delta \alpha_6^i$ between the global $\vec{\Psi}_{6} = \frac{1}{N} \sum_i \psi_{6}(\vec{r}_i)$ and the local $\psi_{6}(\vec{r}_i)$ bond-orientational parameters, $\psi_{6}(\vec{r}_i)\cdot \vec{\Psi}_{6}^* = |\psi_{6}(\vec{r}_i)||\vec{\Psi}_{6}^*|\cos(\Delta \alpha_k^i)$. 
We observe a uniform blue color in the solid and hexatic (Fig.~\ref{fig:highT}(f) \circled{1} and \circled{2}), reflecting the long-range or quasi-long-range of the bond-orientational order.
In the fluid phase, Fig.~\ref{fig:highT}(e) \circled{4}, the snapshot appears almost randomly coloured, due to the short-range of the bond-orientational order. 
In the coexistence of hexatic and fluid phase, Fig.~\ref{fig:highT}(e) \circled{3}, patches with a uniform colour and randomly coloured regions coexist.
\begin{figure}[t]
 \centering
 \includegraphics[angle=0,width=0.48\textwidth]{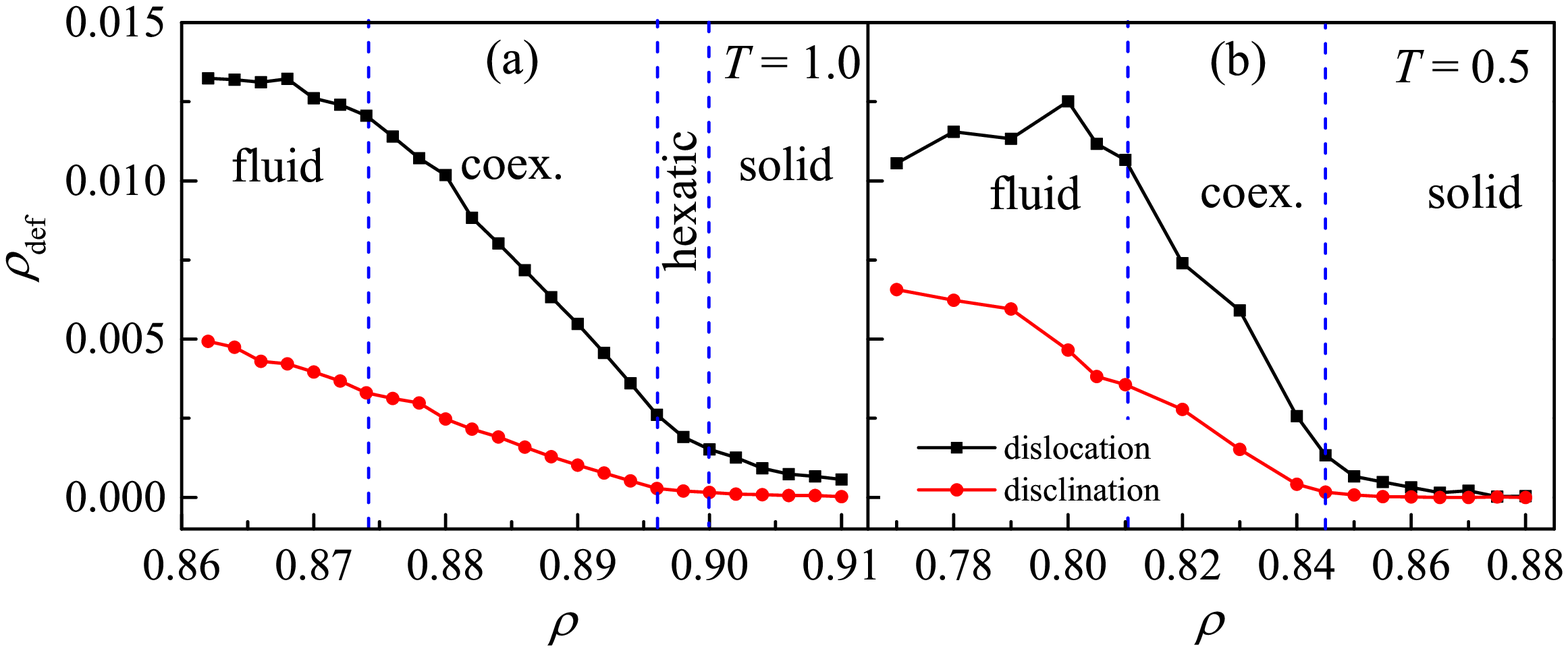}
 \caption{The density dependence of the fraction of isolated dislocations (black squares) and isolated disclinations (red circles) at (a) $T=1.0$ and (b) $T=0.5$. The blue dashed lines separate different phases.
\label{fig:isodef}
}
\end{figure}

All the above observations demonstrate that, at high temperature, when the attractive forces play a minor role, LJ solids melt via a mixed melting scenario. 
This finding agrees with previous investigations of the $1/r^{12}$~\cite{Krauth2015} and WCA~\cite{wcaMelt} systems, but contrasts with ~\cite{Hajibabaei2019} which claimed the hexatic-fluid transition to be continuous. See Ref.~\cite{Li2019} for details.

The high-temperature melting scenario gets modified as the temperature decreases, as we illustrate in Fig.~\ref{fig:lowT}, where we report the equation of state (a) for different temperatures, the translational (b) and bond-orientational  (c) correlation functions, the local density probability distribution (d), and snapshots of the system with particles colour-coded as in Fig.~\ref{fig:highT}.
Melting involves a discontinuous transition at all temperatures, as apparent from the equation of state (a) and the distribution of the local density (d); 
the low-density coexisting phase is, of course, always of liquid type.
The analysis of the correlation functions indicates that the high-density coexisting phase is hexatic for $T \simeq 0.7$ and solid at a lower temperature.
Indeed, only at higher temperatures (e.g., \circled{2} and \circled{6}) the high-density coexisting phase has a exponentially decaying $c(r)$, and a algebraically (or extended) decaying $g_{6}(r)$.
The width of the density range where the hexatic phase occurs, therefore, shrinks as the temperature decreases, until it disappears at $T \simeq 0.7$. 

The direct visualization of the system with particles colour-coded according to $\Delta \alpha_6^i$ supports the above identification of the phases. 
We observe a uniform blue colour in the solid phase (\circled{9}), uncorrelated colours in the fluid phase (\circled{11}), and uncorrelated patches on a blue background in the fluid/solid coexistence region (\circled{10}).

\subsubsection{Topological defects}
The evolution of the density of topological defects across the melting transition further allows distinguishing the different melting scenarios.
Topological defects are particles whose number of nearest neighbours, determined by the Voronoi method, differs from six.
While fluids might have isolated defects, disclinations, in the hexatic phase defects must mainly appear in dislocations ($5-7$ pairs) and in dislocation pairs ($5-7-5-7$ quartets), which do affect the translational order.
In the solid, ideally, no defects are expected.
We investigate the fraction of isolated dislocations and isolated disclinations as the system melts by reducing the density at a high temperature, where the melting is of hard-disk type, and at a low-temperature, where melting occurs through a first-order solid-liquid transition with no hexatic phase.

At high temperature, $T = 1.0$, isolated dislocations start growing on decreasing the density at $\rho=0.90$, which corresponds to the solid-hexatic transition density.
Conversely, the fraction of disclinations starts growing when the system enters the hexatic-fluid coexistence region, at $\rho= 0.896$.
This result is in line with the expected presence~\cite{KT, HN, Y} of a finite fraction of isolated dislocation and very few isolated disclinations, in the hexatic phase. 
In the coexistence phase, the relative amount of fluid-like and hexatic-like particles varies linearly with the density. 
Consistently, the density of dislocations and that of disclinations vary approximately linearly with the density.

At low temperature, $T=0.5$, the fraction of dislocations and that of disclinations start growing at the low-density limit of the solid phase, again supporting the absence of the hexatic phase. 

\begin{figure}[tb]
 \centering
 \includegraphics[angle=0,width=0.48\textwidth]{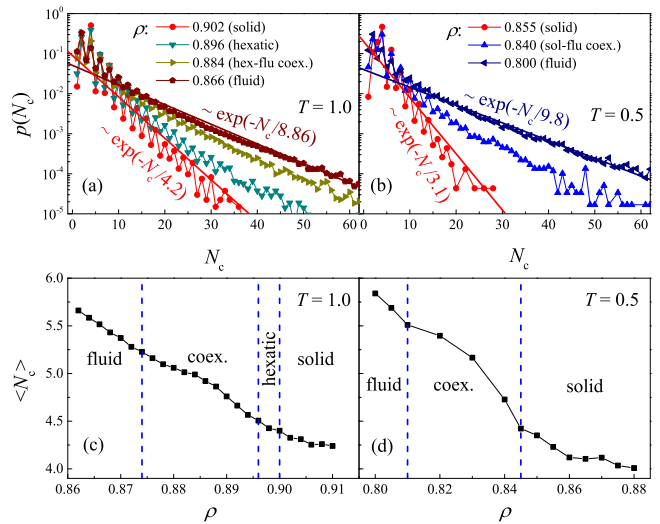}
 \caption{Probability distribution of the size of cluster of defective particles $p(N_{\rm c})$, for selected densities in different phases at (a) $T=1.0$ and (b) $T=0.5$. $p(N_{\rm c})$ shows an exponential decay as indicated by the solid lines in both (a) and (b). 
Panels (c) and (d) illustrate the density dependence of the mean cluster size $\langle N_{\rm c}\rangle$, at $T=1.0$ and $T=0.5$.
\label{fig:clusdef}
}
\end{figure}
Besides occurring isolated and in small clusters, defective particles can also agglomerate in large ones.
Large clusters are not accounted for in the original KTHNY theory~\cite{KT, HN, Y}, but have been observed in systems with power-law interactions~\cite{Krauth2015, digregorio2019clustering}, in hard regular polygons~\cite{Glotzer}, and recently in both passive and active hard disks~\cite{digregorio2019clustering}. 
Here, we evaluate the cluster size probability distribution $p(N_c)$ across the density induced melting transition,
for the temperature values considered in Fig.~\ref{fig:clusdef}.
The cluster size probability distribution decays exponentially, regardless of the phase of the system, both at high and at small temperature, as we illustrate in Fig.~\ref{fig:isodef}(a) and (b), respectively.
This finding contrasts with recent observation in both hard and soft repulsive disks, where the cluster size distribution is power-law close to the hexatic-liquid phase boundary~\cite{digregorio2019clustering}. 
Attractive forces, therefore, influence this property.

While the cluster size distribution is always exponential, the average cluster size $\langle N_{\rm c}\rangle$ increases as the system melts, as we illustrate in Fig.~\ref{fig:isodef}(c) for $T=1.0$, and in Fig.~\ref{fig:isodef}(d) for $T=0.5$. 
In the high-density solid phases $\langle N_{\rm c}\rangle \simeq 4$,  as expected considering that bounded dislocation pairs, which involve $4$ defected particles, dominate in the solid phase. 
As the density decreases, $\langle N_{\rm c}\rangle$ grows significantly in the fluid and coexistence phases. 
These observations do not depend on the existence of the hexatic phase and the melting scenario.

\section{Conclusions\label{sec:end}}
\begin{figure}[htb]
 \centering
 \includegraphics[angle=0,width=0.4\textwidth]{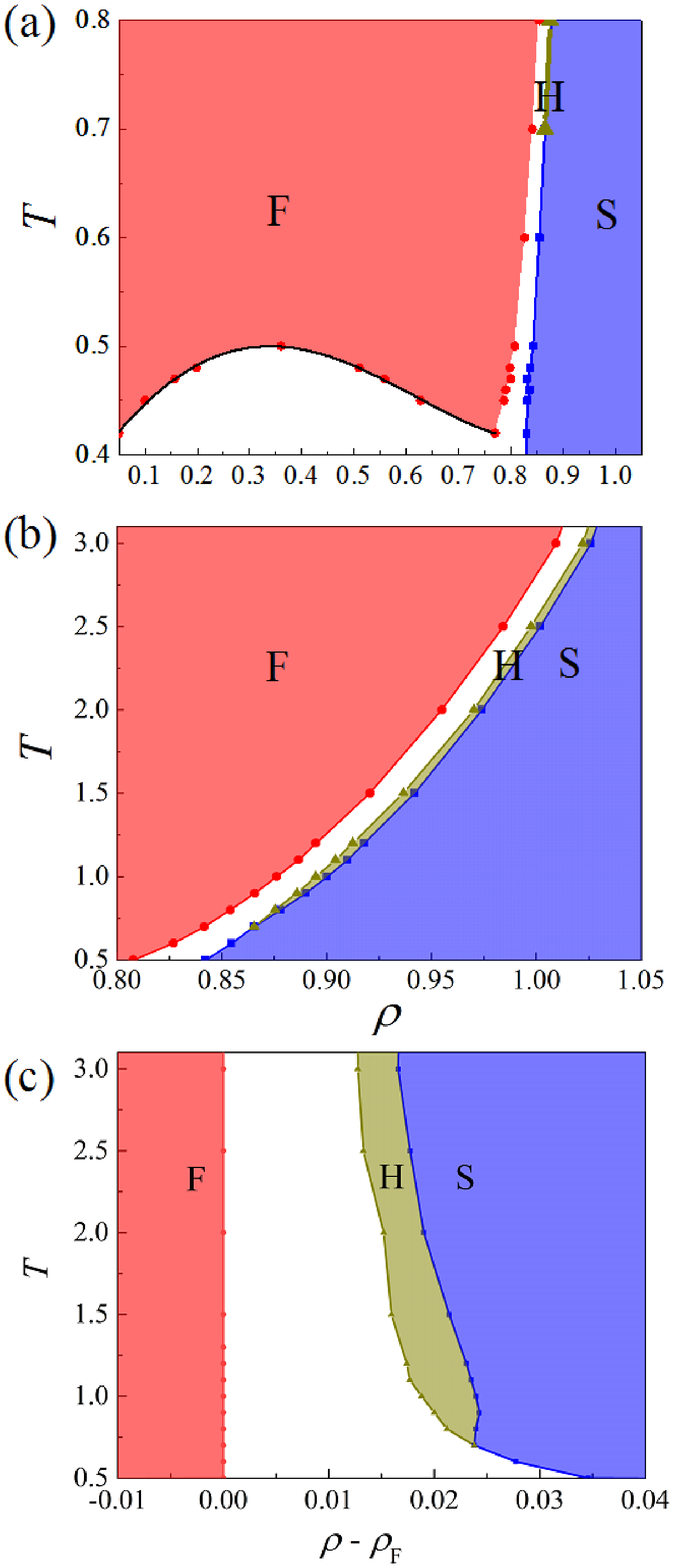}
 \caption{Phase diagram two-dimensional LJ particles.
Panel (a) focuses on the whole $T-\rho$ plane, while panel (b) is a zoom on the high-temperature high-density region where the hexatic phase emerges. In panel (c) the phase diagram is plotted in the $T$ vs $\rho -\rho_{\rm F}$ plane, where $\rho_{\rm F}$ is the coexistence density of the fluid phase.
At each investigated temperature, we mark with symbols the estimated phase boundaries. 
We use colours to distinguish the pure phases, fluid (F), solid (S) and hexatic (H). Coexistence regions, including hexatic-fluid, solid-fluid and liquid-gas coexistence, are white.
\label{fig:pd}
}
\end{figure}

We summarize our findings on the phase behaviour of 2D LJ particles in the phase diagram presented in Fig.~\ref{fig:pd},
which markedly differs from previously speculated one Fig.~\ref{fig:pdNelson}.

Panel (a) illustrates the phase diagram in the temperature range between $0.4$ and $0.8$ to highlight the liquid-gas coexistence region. 
Panel (b) focuses on the high-temperature and high-density region, where the hexatic phase appears. 
Panel (c), which focuses on the same region, illustrates the phase diagram in the $T$ and $\rho-\rho_F$ plane, where $\rho_F$ is the high-density limit of the fluid phase, to highlight the hexatic region. 
The phase diagram reveals two important features. 

First of all, at high temperature, when the attraction in the LJ potential plays a minor role, the system melt via mixed scenario with a continuous solid-hexatic transition and a first-order hexatic-fluid transition.
This melting scenario occurs in the $1/r^{12}$ and WCA systems~\cite{Krauth2015, wcaMelt}. 
In this limit, the density range where the hexatic phase occurs has a constant width.

Secondly, as the temperature decreases the solid-hexatic and the hexatic-fluid transitions densities decrease at different rates so that hexatic density range shrinks.
The hexatic phase disappears $T=0.7$, below which melting occurs via a first-order solid-fluid transition.
We remark that the temperature where the hexatic phase vanishes is much higher than the critical temperature, $T\simeq0.496$.

The proliferation of topological defects drives the melting transition. In agreement with the KTHNY theory, we observed in the hexatic phase isolated dislocations and virtually no isolated disclinations. In the coexistence regions, both kinds of defects are present, and their density varies linearly with the density.
While clusters involving more than a few defective particles are always present, we have found the cluster size distribution to be always exponential, with a characteristic size continuously growing as the system melt. Hence, in this system, melting does not result from a percolation transition of defective particles.

\begin{acknowledgments}
We acknowledge support from the Singapore Ministry of Education
through the Academic Research Fund (Tier 2) MOE2017-T2-1-066 (S) and from the
National Research Foundation Singapore, and are grateful to the National
Supercomputing Centre (NSCC) of Singapore for providing computational resources.
\end{acknowledgments}


\begin{thebibliography}{35}%
\makeatletter
\providecommand \@ifxundefined [1]{%
 \@ifx{#1\undefined}
}%
\providecommand \@ifnum [1]{%
 \ifnum #1\expandafter \@firstoftwo
 \else \expandafter \@secondoftwo
 \fi
}%
\providecommand \@ifx [1]{%
 \ifx #1\expandafter \@firstoftwo
 \else \expandafter \@secondoftwo
 \fi
}%
\providecommand \natexlab [1]{#1}%
\providecommand \enquote  [1]{``#1''}%
\providecommand \bibnamefont  [1]{#1}%
\providecommand \bibfnamefont [1]{#1}%
\providecommand \citenamefont [1]{#1}%
\providecommand \href@noop [0]{\@secondoftwo}%
\providecommand \href [0]{\begingroup \@sanitize@url \@href}%
\providecommand \@href[1]{\@@startlink{#1}\@@href}%
\providecommand \@@href[1]{\endgroup#1\@@endlink}%
\providecommand \@sanitize@url [0]{\catcode `\\12\catcode `\$12\catcode
  `\&12\catcode `\#12\catcode `\^12\catcode `\_12\catcode `\%12\relax}%
\providecommand \@@startlink[1]{}%
\providecommand \@@endlink[0]{}%
\providecommand \url  [0]{\begingroup\@sanitize@url \@url }%
\providecommand \@url [1]{\endgroup\@href {#1}{\urlprefix }}%
\providecommand \urlprefix  [0]{URL }%
\providecommand \Eprint [0]{\href }%
\providecommand \doibase [0]{http://dx.doi.org/}%
\providecommand \selectlanguage [0]{\@gobble}%
\providecommand \bibinfo  [0]{\@secondoftwo}%
\providecommand \bibfield  [0]{\@secondoftwo}%
\providecommand \translation [1]{[#1]}%
\providecommand \BibitemOpen [0]{}%
\providecommand \bibitemStop [0]{}%
\providecommand \bibitemNoStop [0]{.\EOS\space}%
\providecommand \EOS [0]{\spacefactor3000\relax}%
\providecommand \BibitemShut  [1]{\csname bibitem#1\endcsname}%
\let\auto@bib@innerbib\@empty
\bibitem [{\citenamefont {Brinkman}\ \emph {et~al.}(1982)\citenamefont
  {Brinkman}, \citenamefont {Fisher},\ and\ \citenamefont
  {Moncton}}]{Brinkman1982}%
  \BibitemOpen
  \bibfield  {author} {\bibinfo {author} {\bibfnamefont {W.~F.}\ \bibnamefont
  {Brinkman}}, \bibinfo {author} {\bibfnamefont {D.~S.}\ \bibnamefont
  {Fisher}}, \ and\ \bibinfo {author} {\bibfnamefont {D.~E.}\ \bibnamefont
  {Moncton}},\ }\href@noop {} {\bibfield  {journal} {\bibinfo  {journal}
  {Science}\ }\textbf {\bibinfo {volume} {217}},\ \bibinfo {pages} {693}
  (\bibinfo {year} {1982})}\BibitemShut {NoStop}%
\bibitem [{\citenamefont {Heiney}\ \emph {et~al.}(1983)\citenamefont {Heiney},
  \citenamefont {Stephens}, \citenamefont {Birgeneau}, \citenamefont {Horn},\
  and\ \citenamefont {Moncton}}]{Heiney}%
  \BibitemOpen
  \bibfield  {author} {\bibinfo {author} {\bibfnamefont {P.~A.}\ \bibnamefont
  {Heiney}}, \bibinfo {author} {\bibfnamefont {P.~W.}\ \bibnamefont
  {Stephens}}, \bibinfo {author} {\bibfnamefont {R.~J.}\ \bibnamefont
  {Birgeneau}}, \bibinfo {author} {\bibfnamefont {P.~M.}\ \bibnamefont {Horn}},
  \ and\ \bibinfo {author} {\bibfnamefont {D.~E.}\ \bibnamefont {Moncton}},\
  }\href@noop {} {\bibfield  {journal} {\bibinfo  {journal} {Physical Review
  B}\ }\textbf {\bibinfo {volume} {28}},\ \bibinfo {pages} {6416} (\bibinfo
  {year} {1983})},\ \bibinfo {note} {pRB}\BibitemShut {NoStop}%
\bibitem [{\citenamefont {Strandburg}(1988)}]{Strandburg}%
  \BibitemOpen
  \bibfield  {author} {\bibinfo {author} {\bibfnamefont {K.~J.}\ \bibnamefont
  {Strandburg}},\ }\href@noop {} {\bibfield  {journal} {\bibinfo  {journal}
  {Rev. Mod. Phys.}\ }\textbf {\bibinfo {volume} {60}},\ \bibinfo {pages} {161}
  (\bibinfo {year} {1988})}\BibitemShut {NoStop}%
\bibitem [{\citenamefont {Barker}\ \emph {et~al.}(1981)\citenamefont {Barker},
  \citenamefont {Henderson},\ and\ \citenamefont {Abraham}}]{LJpd1981}%
  \BibitemOpen
  \bibfield  {author} {\bibinfo {author} {\bibfnamefont {J.~A.}\ \bibnamefont
  {Barker}}, \bibinfo {author} {\bibfnamefont {D.}~\bibnamefont {Henderson}}, \
  and\ \bibinfo {author} {\bibfnamefont {F.~F.}\ \bibnamefont {Abraham}},\
  }\href@noop {} {\bibfield  {journal} {\bibinfo  {journal} {Physica A}\
  }\textbf {\bibinfo {volume} {106}},\ \bibinfo {pages} {226} (\bibinfo {year}
  {1981})}\BibitemShut {NoStop}%
\bibitem [{\citenamefont {Smit}\ and\ \citenamefont
  {Frenkel}(1991)}]{LJ_vapor_liq}%
  \BibitemOpen
  \bibfield  {author} {\bibinfo {author} {\bibfnamefont {B.}~\bibnamefont
  {Smit}}\ and\ \bibinfo {author} {\bibfnamefont {D.}~\bibnamefont {Frenkel}},\
  }\href@noop {} {\bibfield  {journal} {\bibinfo  {journal} {The Journal of
  Chemical Physics}\ }\textbf {\bibinfo {volume} {94}},\ \bibinfo {pages}
  {5663} (\bibinfo {year} {1991})}\BibitemShut {NoStop}%
\bibitem [{\citenamefont {Singh}\ \emph {et~al.}(1990)\citenamefont {Singh},
  \citenamefont {Pitzer}, \citenamefont {Pablo},\ and\ \citenamefont
  {Prausnitz}}]{Singh90}%
  \BibitemOpen
  \bibfield  {author} {\bibinfo {author} {\bibfnamefont {R.~R.}\ \bibnamefont
  {Singh}}, \bibinfo {author} {\bibfnamefont {K.~S.}\ \bibnamefont {Pitzer}},
  \bibinfo {author} {\bibfnamefont {J.~J.~d.}\ \bibnamefont {Pablo}}, \ and\
  \bibinfo {author} {\bibfnamefont {J.~M.}\ \bibnamefont {Prausnitz}},\
  }\href@noop {} {\bibfield  {journal} {\bibinfo  {journal} {J. Chem. Phys.}\
  }\textbf {\bibinfo {volume} {92}},\ \bibinfo {pages} {5463} (\bibinfo {year}
  {1990})}\BibitemShut {NoStop}%
\bibitem [{\citenamefont {Feng}\ \emph {et~al.}(2000)\citenamefont {Feng},
  \citenamefont {Li},\ and\ \citenamefont {Guo}}]{LJpd_Feng2000}%
  \BibitemOpen
  \bibfield  {author} {\bibinfo {author} {\bibfnamefont {X.}~\bibnamefont
  {Feng}}, \bibinfo {author} {\bibfnamefont {Z.}~\bibnamefont {Li}}, \ and\
  \bibinfo {author} {\bibfnamefont {Z.}~\bibnamefont {Guo}},\ }\href@noop {}
  {\bibfield  {journal} {\bibinfo  {journal} {Chin.Sci.Bull.}\ }\textbf
  {\bibinfo {volume} {45}},\ \bibinfo {pages} {2004} (\bibinfo {year}
  {2000})}\BibitemShut {NoStop}%
\bibitem [{\citenamefont {Phillips}\ \emph {et~al.}(1981)\citenamefont
  {Phillips}, \citenamefont {Bruch},\ and\ \citenamefont
  {Murphy}}]{LJpd_Phillips}%
  \BibitemOpen
  \bibfield  {author} {\bibinfo {author} {\bibfnamefont {J.~M.}\ \bibnamefont
  {Phillips}}, \bibinfo {author} {\bibfnamefont {L.~W.}\ \bibnamefont {Bruch}},
  \ and\ \bibinfo {author} {\bibfnamefont {R.~D.}\ \bibnamefont {Murphy}},\
  }\href@noop {} {\bibfield  {journal} {\bibinfo  {journal} {The Journal of
  Chemical Physics}\ }\textbf {\bibinfo {volume} {75}},\ \bibinfo {pages}
  {5097} (\bibinfo {year} {1981})}\BibitemShut {NoStop}%
\bibitem [{\citenamefont {Shiba}\ \emph {et~al.}(2009)\citenamefont {Shiba},
  \citenamefont {Onuki},\ and\ \citenamefont {Araki}}]{Hayato_EPL}%
  \BibitemOpen
  \bibfield  {author} {\bibinfo {author} {\bibfnamefont {H.}~\bibnamefont
  {Shiba}}, \bibinfo {author} {\bibfnamefont {A.}~\bibnamefont {Onuki}}, \ and\
  \bibinfo {author} {\bibfnamefont {T.}~\bibnamefont {Araki}},\ }\href@noop {}
  {\bibfield  {journal} {\bibinfo  {journal} {Europhys. Lett.}\ }\textbf
  {\bibinfo {volume} {86}},\ \bibinfo {pages} {66004} (\bibinfo {year}
  {2009})}\BibitemShut {NoStop}%
\bibitem [{\citenamefont {Wierschem}\ and\ \citenamefont
  {Manousakis}(2011)}]{Wierschem2011}%
  \BibitemOpen
  \bibfield  {author} {\bibinfo {author} {\bibfnamefont {K.}~\bibnamefont
  {Wierschem}}\ and\ \bibinfo {author} {\bibfnamefont {E.}~\bibnamefont
  {Manousakis}},\ }\href@noop {} {\bibfield  {journal} {\bibinfo  {journal}
  {Phys. Rev. B}\ }\textbf {\bibinfo {volume} {83}},\ \bibinfo {pages} {214108}
  (\bibinfo {year} {2011})}\BibitemShut {NoStop}%
\bibitem [{\citenamefont {Hajibabaei}\ and\ \citenamefont
  {Kim}(2019)}]{Hajibabaei2019}%
  \BibitemOpen
  \bibfield  {author} {\bibinfo {author} {\bibfnamefont {A.}~\bibnamefont
  {Hajibabaei}}\ and\ \bibinfo {author} {\bibfnamefont {K.~S.}\ \bibnamefont
  {Kim}},\ }\href@noop {} {\bibfield  {journal} {\bibinfo  {journal} {Phys.
  Rev. E}\ }\textbf {\bibinfo {volume} {99}},\ \bibinfo {pages} {022145}
  (\bibinfo {year} {2019})}\BibitemShut {NoStop}%
\bibitem [{\citenamefont {Kosterlitz}\ and\ \citenamefont
  {Thouless}(1973)}]{KT}%
  \BibitemOpen
  \bibfield  {author} {\bibinfo {author} {\bibfnamefont {J.~M.}\ \bibnamefont
  {Kosterlitz}}\ and\ \bibinfo {author} {\bibfnamefont {D.~J.}\ \bibnamefont
  {Thouless}},\ }\href@noop {} {\bibfield  {journal} {\bibinfo  {journal} {J.
  Phys. C}\ }\textbf {\bibinfo {volume} {6}},\ \bibinfo {pages} {1181}
  (\bibinfo {year} {1973})}\BibitemShut {NoStop}%
\bibitem [{\citenamefont {Halperin}\ and\ \citenamefont {Nelson}(1978)}]{HN}%
  \BibitemOpen
  \bibfield  {author} {\bibinfo {author} {\bibfnamefont {B.~I.}\ \bibnamefont
  {Halperin}}\ and\ \bibinfo {author} {\bibfnamefont {D.~R.}\ \bibnamefont
  {Nelson}},\ }\href@noop {} {\bibfield  {journal} {\bibinfo  {journal} {Phys.
  Rev. Lett.}\ }\textbf {\bibinfo {volume} {41}},\ \bibinfo {pages} {121}
  (\bibinfo {year} {1978})}\BibitemShut {NoStop}%
\bibitem [{\citenamefont {Young}(1979)}]{Y}%
  \BibitemOpen
  \bibfield  {author} {\bibinfo {author} {\bibfnamefont {A.~P.}\ \bibnamefont
  {Young}},\ }\href@noop {} {\bibfield  {journal} {\bibinfo  {journal} {Phys.
  Rev. B}\ }\textbf {\bibinfo {volume} {19}},\ \bibinfo {pages} {1855}
  (\bibinfo {year} {1979})}\BibitemShut {NoStop}%
\bibitem [{\citenamefont {Bernard}\ and\ \citenamefont
  {Krauth}(2011)}]{Krauth2011}%
  \BibitemOpen
  \bibfield  {author} {\bibinfo {author} {\bibfnamefont {E.~P.}\ \bibnamefont
  {Bernard}}\ and\ \bibinfo {author} {\bibfnamefont {W.}~\bibnamefont
  {Krauth}},\ }\href@noop {} {\bibfield  {journal} {\bibinfo  {journal} {Phys.
  Rev. Lett.}\ }\textbf {\bibinfo {volume} {107}},\ \bibinfo {pages} {155704}
  (\bibinfo {year} {2011})}\BibitemShut {NoStop}%
\bibitem [{\citenamefont {Kapfer}\ and\ \citenamefont
  {Krauth}(2015)}]{Krauth2015}%
  \BibitemOpen
  \bibfield  {author} {\bibinfo {author} {\bibfnamefont {S.~C.}\ \bibnamefont
  {Kapfer}}\ and\ \bibinfo {author} {\bibfnamefont {W.}~\bibnamefont
  {Krauth}},\ }\href@noop {} {\bibfield  {journal} {\bibinfo  {journal} {Phys.
  Rev. Lett.}\ }\textbf {\bibinfo {volume} {114}},\ \bibinfo {pages} {035702}
  (\bibinfo {year} {2015})}\BibitemShut {NoStop}%
\bibitem [{\citenamefont {Li}\ and\ \citenamefont
  {Pica~Ciamarra}(2018)}]{OurPaper}%
  \BibitemOpen
  \bibfield  {author} {\bibinfo {author} {\bibfnamefont {Y.-W.}\ \bibnamefont
  {Li}}\ and\ \bibinfo {author} {\bibfnamefont {M.}~\bibnamefont
  {Pica~Ciamarra}},\ }\href@noop {} {\bibfield  {journal} {\bibinfo  {journal}
  {Phys. Rev. Mater.}\ }\textbf {\bibinfo {volume} {2}},\ \bibinfo {pages}
  {045602} (\bibinfo {year} {2018})}\BibitemShut {NoStop}%
\bibitem [{\citenamefont {Anderson}\ \emph {et~al.}(2017)\citenamefont
  {Anderson}, \citenamefont {Antonaglia}, \citenamefont {Millan}, \citenamefont
  {Engel},\ and\ \citenamefont {Glotzer}}]{Glotzer}%
  \BibitemOpen
  \bibfield  {author} {\bibinfo {author} {\bibfnamefont {J.~A.}\ \bibnamefont
  {Anderson}}, \bibinfo {author} {\bibfnamefont {J.}~\bibnamefont
  {Antonaglia}}, \bibinfo {author} {\bibfnamefont {J.~A.}\ \bibnamefont
  {Millan}}, \bibinfo {author} {\bibfnamefont {M.}~\bibnamefont {Engel}}, \
  and\ \bibinfo {author} {\bibfnamefont {S.~C.}\ \bibnamefont {Glotzer}},\
  }\href@noop {} {\bibfield  {journal} {\bibinfo  {journal} {Phys. Rev. X}\
  }\textbf {\bibinfo {volume} {7}},\ \bibinfo {pages} {021001} (\bibinfo {year}
  {2017})}\BibitemShut {NoStop}%
\bibitem [{\citenamefont {Zu}\ \emph {et~al.}(2016)\citenamefont {Zu},
  \citenamefont {Liu}, \citenamefont {Tong},\ and\ \citenamefont
  {Xu}}]{ningxu}%
  \BibitemOpen
  \bibfield  {author} {\bibinfo {author} {\bibfnamefont {M.}~\bibnamefont
  {Zu}}, \bibinfo {author} {\bibfnamefont {J.}~\bibnamefont {Liu}}, \bibinfo
  {author} {\bibfnamefont {H.}~\bibnamefont {Tong}}, \ and\ \bibinfo {author}
  {\bibfnamefont {N.}~\bibnamefont {Xu}},\ }\href@noop {} {\bibfield  {journal}
  {\bibinfo  {journal} {Phys. Rev. Lett.}\ }\textbf {\bibinfo {volume} {117}},\
  \bibinfo {pages} {85702} (\bibinfo {year} {2016})}\BibitemShut {NoStop}%
\bibitem [{\citenamefont {Russo}\ and\ \citenamefont
  {Wilding}(2017)}]{John_Russo}%
  \BibitemOpen
  \bibfield  {author} {\bibinfo {author} {\bibfnamefont {J.}~\bibnamefont
  {Russo}}\ and\ \bibinfo {author} {\bibfnamefont {N.~B.}\ \bibnamefont
  {Wilding}},\ }\href@noop {} {\bibfield  {journal} {\bibinfo  {journal} {Phys.
  Rev. Lett.}\ }\textbf {\bibinfo {volume} {119}},\ \bibinfo {pages} {115702}
  (\bibinfo {year} {2017})}\BibitemShut {NoStop}%
\bibitem [{\citenamefont {Thorneywork}\ \emph {et~al.}(2017)\citenamefont
  {Thorneywork}, \citenamefont {Abbott}, \citenamefont {Aarts},\ and\
  \citenamefont {Dullens}}]{Experiment_harddisc}%
  \BibitemOpen
  \bibfield  {author} {\bibinfo {author} {\bibfnamefont {A.~L.}\ \bibnamefont
  {Thorneywork}}, \bibinfo {author} {\bibfnamefont {J.~L.}\ \bibnamefont
  {Abbott}}, \bibinfo {author} {\bibfnamefont {D.~G. A.~L.}\ \bibnamefont
  {Aarts}}, \ and\ \bibinfo {author} {\bibfnamefont {R.~P.~A.}\ \bibnamefont
  {Dullens}},\ }\href@noop {} {\bibfield  {journal} {\bibinfo  {journal} {Phys.
  Rev. Lett.}\ }\textbf {\bibinfo {volume} {118}},\ \bibinfo {pages} {158001}
  (\bibinfo {year} {2017})}\BibitemShut {NoStop}%
\bibitem [{\citenamefont {Li}\ and\ \citenamefont
  {Ciamarra}(2020)}]{Massimo_2020PRLmelting}%
  \BibitemOpen
  \bibfield  {author} {\bibinfo {author} {\bibfnamefont {Y.-W.}\ \bibnamefont
  {Li}}\ and\ \bibinfo {author} {\bibfnamefont {M.~P.}\ \bibnamefont
  {Ciamarra}},\ }\href@noop {} {\bibfield  {journal} {\bibinfo  {journal}
  {Phys. Rev. Lett.}\ }\textbf {\bibinfo {volume} {124}},\ \bibinfo {pages}
  {218002} (\bibinfo {year} {2020})}\BibitemShut {NoStop}%
\bibitem [{\citenamefont {Nelson}(2002)}]{Nelson_book}%
  \BibitemOpen
  \bibfield  {author} {\bibinfo {author} {\bibfnamefont {D.~R.}\ \bibnamefont
  {Nelson}},\ }\href@noop {} {\emph {\bibinfo {title} {Defects and Geometry in
  Condensed Matter Physics}}}\ (\bibinfo  {publisher} {Cambridge University
  Press, Cambridge},\ \bibinfo {year} {2002})\BibitemShut {NoStop}%
\bibitem [{\citenamefont {Nelson}\ and\ \citenamefont
  {Halperin}(1979)}]{Nelson1979}%
  \BibitemOpen
  \bibfield  {author} {\bibinfo {author} {\bibfnamefont {D.~R.}\ \bibnamefont
  {Nelson}}\ and\ \bibinfo {author} {\bibfnamefont {B.~I.}\ \bibnamefont
  {Halperin}},\ }\href@noop {} {\bibfield  {journal} {\bibinfo  {journal}
  {Phys. Rev. B}\ }\textbf {\bibinfo {volume} {19}},\ \bibinfo {pages} {2457}
  (\bibinfo {year} {1979})}\BibitemShut {NoStop}%
\bibitem [{\citenamefont {Weeks}\ \emph {et~al.}(1971)\citenamefont {Weeks},
  \citenamefont {Chandler},\ and\ \citenamefont {Andersen}}]{WCA}%
  \BibitemOpen
  \bibfield  {author} {\bibinfo {author} {\bibfnamefont {J.~D.}\ \bibnamefont
  {Weeks}}, \bibinfo {author} {\bibfnamefont {D.}~\bibnamefont {Chandler}}, \
  and\ \bibinfo {author} {\bibfnamefont {H.~C.}\ \bibnamefont {Andersen}},\
  }\href@noop {} {\bibfield  {journal} {\bibinfo  {journal} {The Journal of
  Chemical Physics}\ }\textbf {\bibinfo {volume} {54}},\ \bibinfo {pages}
  {5237} (\bibinfo {year} {1971})}\BibitemShut {NoStop}%
\bibitem [{\citenamefont {Khali}\ \emph {et~al.}(2020)\citenamefont {Khali},
  \citenamefont {Chakraborty},\ and\ \citenamefont {Chaudhuri}}]{wcaMelt}%
  \BibitemOpen
  \bibfield  {author} {\bibinfo {author} {\bibfnamefont {S.~S.}\ \bibnamefont
  {Khali}}, \bibinfo {author} {\bibfnamefont {D.}~\bibnamefont {Chakraborty}},
  \ and\ \bibinfo {author} {\bibfnamefont {D.}~\bibnamefont {Chaudhuri}},\
  }\href@noop {} {\bibfield  {journal} {\bibinfo  {journal} {arXiv:2007.00297}\
  } (\bibinfo {year} {2020})}\BibitemShut {NoStop}%
\bibitem [{\citenamefont {Du}\ \emph {et~al.}(2017)\citenamefont {Du},
  \citenamefont {Doxastakis}, \citenamefont {Hilou},\ and\ \citenamefont
  {Biswal}}]{Di_SoftMatter}%
  \BibitemOpen
  \bibfield  {author} {\bibinfo {author} {\bibfnamefont {D.}~\bibnamefont
  {Du}}, \bibinfo {author} {\bibfnamefont {M.}~\bibnamefont {Doxastakis}},
  \bibinfo {author} {\bibfnamefont {E.}~\bibnamefont {Hilou}}, \ and\ \bibinfo
  {author} {\bibfnamefont {S.~L.}\ \bibnamefont {Biswal}},\ }\href@noop {}
  {\bibfield  {journal} {\bibinfo  {journal} {Soft Matter}\ }\textbf {\bibinfo
  {volume} {13}},\ \bibinfo {pages} {1548} (\bibinfo {year}
  {2017})}\BibitemShut {NoStop}%
\bibitem [{\citenamefont {Li}\ \emph {et~al.}(2019)\citenamefont {Li},
  \citenamefont {Xiao}, \citenamefont {Wang}, \citenamefont {Wen},\ and\
  \citenamefont {Wang}}]{Bo_Prx}%
  \BibitemOpen
  \bibfield  {author} {\bibinfo {author} {\bibfnamefont {B.}~\bibnamefont
  {Li}}, \bibinfo {author} {\bibfnamefont {X.}~\bibnamefont {Xiao}}, \bibinfo
  {author} {\bibfnamefont {S.}~\bibnamefont {Wang}}, \bibinfo {author}
  {\bibfnamefont {W.}~\bibnamefont {Wen}}, \ and\ \bibinfo {author}
  {\bibfnamefont {Z.}~\bibnamefont {Wang}},\ }\href@noop {} {\bibfield
  {journal} {\bibinfo  {journal} {Phys. Rev. X}\ }\textbf {\bibinfo {volume}
  {9}},\ \bibinfo {pages} {031032} (\bibinfo {year} {2019})}\BibitemShut
  {NoStop}%
\bibitem [{\citenamefont {Allen}(1987)}]{Allen_book}%
  \BibitemOpen
  \bibfield  {author} {\bibinfo {author} {\bibfnamefont {M.}~\bibnamefont
  {Allen}},\ }\href@noop {} {\emph {\bibinfo {title} {Computer Simulation of
  Liquids}}}\ (\bibinfo  {publisher} {Oxford University Press, Oxford},\
  \bibinfo {year} {1987})\BibitemShut {NoStop}%
\bibitem [{\citenamefont {Zhu}\ \emph {et~al.}(2013)\citenamefont {Zhu},
  \citenamefont {Liu}, \citenamefont {Li}, \citenamefont {Qian}, \citenamefont
  {Milano},\ and\ \citenamefont {Lu}}]{Galamost}%
  \BibitemOpen
  \bibfield  {author} {\bibinfo {author} {\bibfnamefont {Y.}~\bibnamefont
  {Zhu}}, \bibinfo {author} {\bibfnamefont {H.}~\bibnamefont {Liu}}, \bibinfo
  {author} {\bibfnamefont {Z.}~\bibnamefont {Li}}, \bibinfo {author}
  {\bibfnamefont {H.}~\bibnamefont {Qian}}, \bibinfo {author} {\bibfnamefont
  {G.}~\bibnamefont {Milano}}, \ and\ \bibinfo {author} {\bibfnamefont
  {Z.}~\bibnamefont {Lu}},\ }\href@noop {} {\bibfield  {journal} {\bibinfo
  {journal} {J. Comput. Chem.}\ }\textbf {\bibinfo {volume} {34}},\ \bibinfo
  {pages} {2197} (\bibinfo {year} {2013})}\BibitemShut {NoStop}%
\bibitem [{\citenamefont {Li}\ and\ \citenamefont
  {Ciamarra}(2019{\natexlab{a}})}]{Massimo_PRE2019}%
  \BibitemOpen
  \bibfield  {author} {\bibinfo {author} {\bibfnamefont {Y.-W.}\ \bibnamefont
  {Li}}\ and\ \bibinfo {author} {\bibfnamefont {M.~P.}\ \bibnamefont
  {Ciamarra}},\ }\href@noop {} {\bibfield  {journal} {\bibinfo  {journal}
  {Phys. Rev. E}\ }\textbf {\bibinfo {volume} {100}},\ \bibinfo {pages}
  {062606} (\bibinfo {year} {2019}{\natexlab{a}})}\BibitemShut {NoStop}%
\bibitem [{\citenamefont {Mermin}\ and\ \citenamefont {Wagner}(1966)}]{Mermin}%
  \BibitemOpen
  \bibfield  {author} {\bibinfo {author} {\bibfnamefont {N.~D.}\ \bibnamefont
  {Mermin}}\ and\ \bibinfo {author} {\bibfnamefont {H.}~\bibnamefont
  {Wagner}},\ }\href@noop {} {\bibfield  {journal} {\bibinfo  {journal} {Phys.
  Rev. Lett.}\ }\textbf {\bibinfo {volume} {17}},\ \bibinfo {pages} {1133}
  (\bibinfo {year} {1966})}\BibitemShut {NoStop}%
\bibitem [{\citenamefont {Schrader}\ \emph {et~al.}(2009)\citenamefont
  {Schrader}, \citenamefont {Virnau},\ and\ \citenamefont {Binder}}]{Schrader}%
  \BibitemOpen
  \bibfield  {author} {\bibinfo {author} {\bibfnamefont {M.}~\bibnamefont
  {Schrader}}, \bibinfo {author} {\bibfnamefont {P.}~\bibnamefont {Virnau}}, \
  and\ \bibinfo {author} {\bibfnamefont {K.}~\bibnamefont {Binder}},\
  }\href@noop {} {\bibfield  {journal} {\bibinfo  {journal} {Phys. Rev. E}\
  }\textbf {\bibinfo {volume} {79}},\ \bibinfo {pages} {061104} (\bibinfo
  {year} {2009})}\BibitemShut {NoStop}%
\bibitem [{\citenamefont {Li}\ and\ \citenamefont
  {Ciamarra}(2019{\natexlab{b}})}]{Li2019}%
  \BibitemOpen
  \bibfield  {author} {\bibinfo {author} {\bibfnamefont {Y.-W.}\ \bibnamefont
  {Li}}\ and\ \bibinfo {author} {\bibfnamefont {M.~P.}\ \bibnamefont
  {Ciamarra}},\ }\href@noop {} {\bibfield  {journal} {\bibinfo  {journal}
  {Physical Review E}\ }\textbf {\bibinfo {volume} {100}},\ \bibinfo {pages}
  {062606} (\bibinfo {year} {2019}{\natexlab{b}})}\BibitemShut {NoStop}%
\bibitem [{\citenamefont {Digregorio}\ \emph {et~al.}(2019)\citenamefont
  {Digregorio}, \citenamefont {Levis}, \citenamefont {Cugliandolo},
  \citenamefont {Gonnella},\ and\ \citenamefont
  {Pagonabarraga}}]{digregorio2019clustering}%
  \BibitemOpen
  \bibfield  {author} {\bibinfo {author} {\bibfnamefont {P.}~\bibnamefont
  {Digregorio}}, \bibinfo {author} {\bibfnamefont {D.}~\bibnamefont {Levis}},
  \bibinfo {author} {\bibfnamefont {L.~F.}\ \bibnamefont {Cugliandolo}},
  \bibinfo {author} {\bibfnamefont {G.}~\bibnamefont {Gonnella}}, \ and\
  \bibinfo {author} {\bibfnamefont {I.}~\bibnamefont {Pagonabarraga}},\
  }\href@noop {} {\  (\bibinfo {year} {2019})},\ \Eprint
  {http://arxiv.org/abs/1911.06366} {arXiv:1911.06366} \BibitemShut {NoStop}%
\end{thebibliography}

%
\end{document}